\documentclass[conference]{IEEEtran}
\usepackage{verbatim} 
\usepackage{graphicx}
\usepackage{subfigure}
\usepackage{times}
\usepackage{caption}
\usepackage{epsfig}
\usepackage{stfloats}
\usepackage[noadjust]{cite}
\usepackage{enumerate}
\usepackage{amssymb}
\usepackage{multirow}
\usepackage{latexsym}
\usepackage{color}
\usepackage{amsmath}
\usepackage[noend]{algpseudocode}
\usepackage{epstopdf}
\usepackage{tikz}
\usepackage{algorithm}
\usepackage{mathtools}

\allowdisplaybreaks
\usepackage{etoolbox}

 \makeatletter 
 \def\@eqnnum{{\normalsize \normalcolor (\theequation)}} 
  \makeatother
\usepackage{etex}
\usepackage{tabularx} 
\usepackage{algorithm}
\usepackage{graphicx}
\usepackage{algpseudocode}
\usepackage{tikz}
\usetikzlibrary{shapes,arrows}
\usepackage{pstricks}
\usepackage{pst-node,pst-blur}
\usetikzlibrary{arrows}
\usepackage{amsfonts}
\usepackage{bm}
\usepackage{tabularx}
\usepackage{enumerate}
\usepackage{subfigure}
\usepackage{amsmath}
\usepackage{amssymb}
\usepackage{booktabs} 
\usepackage{multirow} 
\usepackage{mathtools} 
\usepackage{epsfig}
\usepackage{epstopdf}
\usepackage{array}

\usepackage[noadjust]{cite}
\usepackage{caption}
\usepackage{etoolbox}

\allowdisplaybreaks
\makeatletter
\def\set@curr@file#1{%
  \begingroup
    \escapechar\m@ne
    \xdef\@curr@file{\expandafter\string\csname #1\endcsname}%
  \endgroup
}
\def\quote@name#1{"\quote@@name#1\@gobble""}
\def\quote@@name#1"{#1\quote@@name}
\def\unquote@name#1{\quote@@name#1\@gobble"}
\makeatother
\usepackage{graphics}

\begin{document}

\title{Circuit Characterization of IRS to Control Beamforming Design for Efficient Wireless Communication}
\author{\IEEEauthorblockN{Bhupendra Sharma\IEEEauthorrefmark{1}, Anirudh Agarwal\IEEEauthorrefmark{2}, Deepak Mishra\IEEEauthorrefmark{3} and Soumitra Debnath\IEEEauthorrefmark{4}}
\IEEEauthorblockA{\IEEEauthorrefmark{1}\IEEEauthorrefmark{2}\IEEEauthorrefmark{4}Department of ECE, 
The LNM Institute of Information Technology, Jaipur, India\\
\IEEEauthorrefmark{3}School of Electrical Engineering and Telecommunications, University of New South Wales, Sydney, Australia
\\E-mails: bhupendrasharma.y19@lnmiit.ac.in\IEEEauthorrefmark{1}, anirudh.agarwal@lnmiit.ac.in\IEEEauthorrefmark{2}, d.mishra@unsw.edu.au\IEEEauthorrefmark{3}, soumitra@lnmiit.ac.in\IEEEauthorrefmark{4}
}}




\maketitle

\begin{abstract}
  Intelligent reflecting surface (IRS) has emerged as a transforming solution to enrich wireless communications by efficiently reconfiguring the propagation environment. In this paper, a novel IRS circuit characterization model is proposed for practical beamforming design incorporating various electrical parameters of the meta-surface unit cell. Specifically, we have modelled the IRS control parameters, phase shift (PS) and reflection amplitude (RA) at the communication receiver, in addition to the circuit level parameter, variable effective capacitance $C$ of IRS unit cell. We have obtained closed-form expressions of PS, RA and $C$ in terms of transmission frequency of signal incident to IRS and various electrical parameters of IRS circuit, with a novel touch towards an accurate analytical model for a better beamforming design perspective. Numerical results demonstrate the efficacy of the proposed characterization thereby providing key design insights.
\end{abstract}


\section{INTRODUCTION}
Intelligent reflecting surface (IRS) is believed to be a promising technology for an accurate beam orientation towards users at different locations, by effectively controlling the phase shift (PS) and reflection amplitude (RA) of reflected signals. An IRS consists of an array of IRS units that reflect the incoming signal independently at the desired PS and RA. IRS has been ubiquitously used in various advanced fields of next generation wireless communications including beyond-5G and 6G networks. It is due its easier integration into side-walls, ceilings or any flat surface in the vicinity of communication transmitter thereby guiding the incoming signals in different directions. However, it is important to accurately characterize the IRS unit circuitry for an efficient beamforming design (BD) with controlled wireless signal propagation accommodating the effect of various electrical parameters like equivalent IRS capacitance, inductance, range of incident signal frequency, in addition to various IRS material properties.

\subsection{State-of-the-Art}
Several existing IRS related works have envisioned IRS as the potential candidate to support the sixth-generation wireless platforms \cite{sixth_gen_wn}. IRS has shown promising advantages and applications in \cite{sixth_gen_app}, including integration of IRS technology with wireless systems in sufficient depth. Consequently, the underlying problems of resource allocation and capacity analysis have been optimally solved in the context of IRS-aided wireless communications \cite{d_mishra, e_basar, m_renzo}. In \cite{d_mishra}, channel estimation was investigated with BD for IRS-assisted multi-antenna wireless energy transfer. In \cite{pow_scal_law}, authors proposed different IRS power scaling laws for massive multiple-input multiple-output (MIMO)-based wireless transmission. In \cite{m_renzo}, the role of artificial intelligence was underlined for smart radio environments based on IRS. 
 
Authors in \cite{stat_csi} have proposed analytical expressions for optimal up-link achievable rate with random number of base station antennas in an IRS-aided wireless communication system (IWCS), thereby discussing various power scaling laws and optimizing the PS at IRS. Moreover, they have also highlighted the key benefits of adjoining IRS into conventional massive MIMO wireless systems via rigorous computer simulations. In \cite{PSO}, authors optimized the transmit passive beamforming using two IRS-based system using particle swarm optimization algorithm. They also simulated the maximum signal power at receiver end. 

Few researchers have also worked on design and modeling of IRS for an enhanced IWCS. In \cite{samith}, phase-dependent RA variation has been derived to characterize the fundamental relationship between PS and RA for beamforming optimization. \cite{physics} have uniquely researched upon the physics of IRS-assisted wireless communication with major focus on the pathloss modeling. Authors in \cite{mirza} have considered constant RA for proposing a PS design model for IRS system with channel estimation. More recently, \cite{cai} have studied the PS-RA-frequency relationship of reflected signal and modelled the reflection coefficient (RC) for an IWCS. 

Author in \cite{emil_b} introduced a two dimensional array structure of a metamaterial cell that controls the behaviour of an electromagnetic wave by tuning the impedance of the IRS. This impedance was was controlled using a switch. It was found that by turning the switch position, RC can be altered, which further varies the PS and RA of the reflected signal. In \cite{MIMO}, authors proposed an IRS architecture analysis to achieve an amplitude and phase varying modulation which enables the design for MIMO-based quadrature amplitude modulation in wireless transmission.

\subsection{Research Gap and Motivation}
Most of the existing IRS-assisted BD works \cite{samith, physics, mirza} in IWCS have either considered PS or PS-dependent RA for their modeling purposes. Moreover, above models do not incorporate the effect of all the circuit level electrical parameters of the IRS unit cell. Liu et al. \cite{Liu} did incorporate PS as well as RA in terms of the loss resistance and equivalent capacitance of an IRS unit, but they did not derive the closed-form expressions of PS, RA and effective capacitance for design and analysis of the IRS model. 

\subsection{Novelty and Scope of the Work}
In this work, we introduce a unique IRS model for BD to enable efficient wireless communication. This is achieved by specifically controlling the passive PS as well as RA as a function of incident frequency (IF) and other physical properties of IRS. For design perspective, we also derive a closed-form solution of IRS capacitance in terms of PS, and RA. \textit{To the best of our knowledge, this is the first work that relates the circuit level electrical parameters with IRS control parameters PS and RA along with the impact of IF for BD in wireless communications}. 

 \subsection{Major Contributions of the Work}
The key contributions of this work are four-fold, which are summarised as follows:
\begin{enumerate}
    \item A unique IRS circuit characterization is proposed for BD, in terms of phase dependent parameters like PS and RA at the receiving end of WCS.
    \item Closed-form expressions are derived for PS as well as RA in terms of IF and other IRS electrical parameters.
    \item Closed-form solutions are also obtained for IRS equivalent capacitance for design perspective.
    \item Numerical results with key design insights are provided to prove the efficacy of our proposed model.
\end{enumerate}

\section{System Description and Existing IRS Model}
In this Section, we first describe the IRS system model including its structural layout and dimensional architecture. Further, the details of the existing IRS design model are provided.
\subsection{System Description}
\begin{figure}
    \centering
    \includegraphics[width=3.48in, height=2.8in]{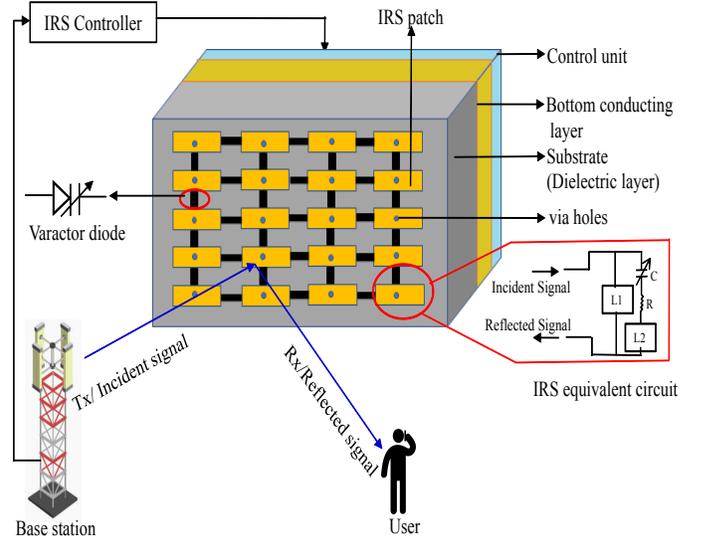}
    \caption{IRS  system model structural layout}
    \label{fig1}
 \end{figure}

The physical structure of the IRS model in Fig. \ref{fig1} illustrates IRS-aided communication between a base station as transmitter (Tx) and a single antenna mobile user as receiver (Rx). In IRS, the metamaterial layer acts as a reflecting surface which can shift the direction of reflected wave, thereby establishing Tx-Rx communication link by following generalized Snell’s law \cite{physics}. In general an IRS, assembled on a printed circuit board (PCB), is composed of multiple reflecting elements with a periodic arrangement of rectangular metal patches on top layer of PCB in addition to a metal sheet on its bottom layer. Both the layers are connected via a varactor diode whose capacitance can be varied by adjusting the applied DC bias voltage. In the same figure, an equivalent parallel resonant circuit is also depicted which illustrates the electrical characteristics of the IRS element. For controlling the bias voltage, a smart IRS controller is used which connects transmitting base station with the IRS control segment using a separate wireless link. 

\begin{figure}
    \centering
     \includegraphics[width=3.48in, height=2.2in]{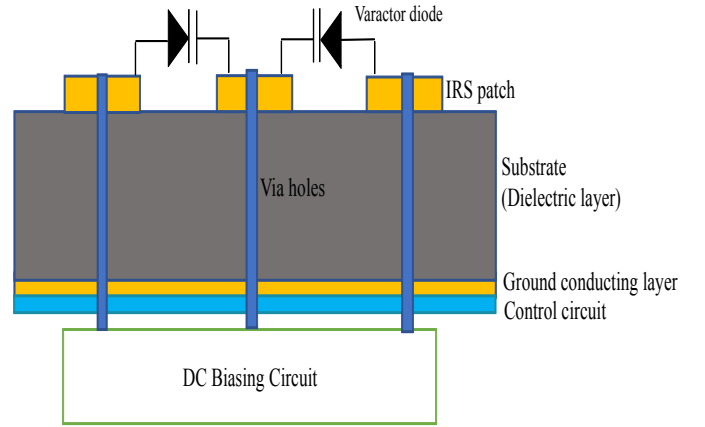} 
     \caption{Dimensional architecture of IRS }
     \label{figure2}
\end{figure}

Furthermore, the dimensional architecture of an IRS element is depicted in Fig. \ref{figure2}. In this design architecture, a tunable capacitor is used in each IRS reflector unit cell and desired PS is achieved by adjusting the DC bias voltage applied to the varactor diode via holes.Via hole is use to provide a conductive path for signal from one layer to another layer.
 
In particular, by tuning equivalent capacitance $C$, desired PS and RA can be achieved which are significant in an IRS design model which can actually enhance the capacity of an IWCS. However, PS and RA also depend on other important parameters related to the physical property of metamaterial, such as bottom layer inductance $L_1$, top layer inductance $L_2$ of IRS architecture and loss resistance $R$ of equivalent circuit.

\subsection{Existing PS and RA based IRS Model}
The main objective of modeling an IRS unit is to maximize the reflection of a wave that is directly incident on IRS. When IRS unit impedance gets mismatched then reflection occurs. This is governed by RC which indicates the fraction of wave reflected due to the impedance mismatching between impedance of the metamaterial-made IRS unit cell $Z$ and its characteristic impedance $Z_0$. In \cite{samith}, authors proposed and experimentally validated a practical IRS model that represents the relation of phase-dependent RA variation. For accomplishing this task, they considered RC $\Gamma_{rc}$ of an IRS element in terms of $C$ and $R$, that is expressed as,
\vspace{0.25cm}
\begin{equation}\label{rcz}
  \Gamma_{rc}(C,R)  = \frac{Z(C,R) - Z_0}{Z(C,R) + Z_0},
\end{equation}

 where $Z(C,R)$ denotes $Z$ as a function of $C$ and $R$. Similarly in \cite{cai}, $\Gamma_{rc}$ is represented in terms of different electrical parameters of equivalent circuit of IRS element and IF as:
 \vspace{0.25cm}
\begin{equation}\label{frcz}
  \Gamma_{rc}(C,f,R,L_1,L_2)  = \frac{Z(C,f,R,L_1,L_2) - Z_0}{Z(C,f,R,L_1,L_2,) + Z_0},
\end{equation}

where $Z(C,f,R,L_1,L_2)$ is a function of $C$, $f$, $R$, $L_1$ and $L_2$, which is mathematically defined as,

\begin{equation}\label{alpha}
    Z(C,f,R,L_1,L_2) = \frac{j\omega L_1(j\omega L_2+\frac{1}{j\omega C}+R)}{j\omega L_1+(j\omega L_2+\frac{1}{j\omega C}+R)} \triangleq \mathcal{Z},
\end{equation}
 where $\omega$ denotes the angular frequency of incident signal.
\section{Proposed IRS Circuit Characterization for BD} 
 We next propose a unique analytical model of IRS for BD in a WCS. Specifically, we intend to obtain relations of PS $\theta$ and RA $\mathcal{R}$ of the signal reflected by IRS with IF of transmitted signal, effective impedance of the IRS element and other physical properties of IRS. Further, particularly for designer's point of interest, we also provide closed-form expressions for equivalent capacitance in terms of PS as well as RA. These circuit relationships would help in controlling IRS-aided BD by selecting appropriate PS and RA as per requirement. For this purpose, an IRS acting as a passive reflector \cite{e_basar}, we first consider that signal with frequency $f$ coming from the communication source towards IRS as incident signal, and the signal received at the destination from IRS as the reflected signal. Then, we characterize the fundamental relationship of RC with PS and RA as \cite{samith, cai},
 \vspace{0.25cm}
\begin{equation}\label{rcoff}
  \Gamma_{rc}(C,f,R,L_1,L_2) = \mathcal{R} e^{j\theta},
\end{equation}
\noindent  where $\Gamma_{rc}$ is a complex quantity with magnitude $\mathcal{R}\in [0,1]$ and phase $\theta \in [-\pi, \pi]$.

\subsection{PS-based IRS Circuit Characterization}
Here we provide PS assisted model for IRS-induced BD. Consequently, utilizing \eqref{frcz} and \eqref{alpha} in \eqref{rcoff} results in,

\begin{equation} \label{alphaz}
  e^{j\theta} = \frac{1}{\mathcal{R}}\left(\frac{\mathcal{Z}-Z_0}{\mathcal{Z}+Z_0}\right).
\end{equation}
Using \eqref{alpha}, \eqref{alphaz} can be further simplified as,
\vspace{0.25cm}
  \begin{equation}\label{ej}
     e^{j\theta} = \frac{\frac{(-\omega^2 L_1 L_2+ \frac{L_1}{C}-RZ_0)+j(\omega RL_1-Z_0\omega L_1-Z_0 \omega L_2+\frac{Z_0}{\omega C} )}{(-\omega^2 L_1 L_2+ \frac{L_1}{C}+RZ_0)+j(\omega RL_1+Z_0\omega L_1+Z_0 \omega L_2-\frac{Z_0}{\omega C} )}}{\mathcal{R}}.
 \end{equation}
With Euler's form, \eqref{ej} can be written as,
\vspace{0.25cm}
  \begin{equation}\label{real_img}
     \cos\theta+j \sin\theta = \frac{AC\mathcal{R}+BD\mathcal{R}}{(C\mathcal{R})^2+(D\mathcal{R})^2}+j\frac{BC\mathcal{R}-AD\mathcal{R}}{(C\mathcal{R})^2+(D\mathcal{R})^2},
  \end{equation}

where $\mathcal{A}\triangleq-\omega^2 L_1 L_2+ \frac{L_1}{C}-RZ_0$,
$\mathcal{B}\triangleq\omega RL_1-Z_0\omega L_1-Z_0 \omega L_2+\frac{Z_0}{\omega C}$, $\mathcal{C}\triangleq-\omega^2 L_1 L_2+ \frac{L_1}{C}+RZ_0$ and $\mathcal{D}\triangleq\omega RL_1+Z_0\omega L_1+Z_0 \omega L_2-\frac{Z_0}{\omega C}$.\\
Now, equating real and imaginary parts in \eqref{real_img} provides,
\vspace{0.25cm}
 \begin{equation} \label{cos}
     \cos\theta = \frac{\mathcal{A}\mathcal{C}\mathcal{R}+\mathcal{B}\mathcal{D}\mathcal{R}}{(\mathcal{C}\mathcal{R})^2+(\mathcal{D}\mathcal{R})^2},
 \end{equation}
 
 \begin{equation}\label{sin}
     \sin\theta =\frac{\mathcal{B}\mathcal{C}\mathcal{R}-\mathcal{A}\mathcal{D}\mathcal{R}}{(\mathcal{C}\mathcal{R})^2+(\mathcal{D}\mathcal{R})^2},
 \end{equation}
 Therefore from \eqref{cos} and \eqref{sin}, PS can be obtained as,
 
 \begin{equation} \label{theta}
    \theta=\tan^{-1}\left(\frac{\mathcal{B}\mathcal{C}-\mathcal{A}\mathcal{D}}{\mathcal{A}\mathcal{C}+\mathcal{B}\mathcal{D}}\right) \triangleq \theta(C,f,R,L_1,L_2).
 \end{equation}

Thus a unique relation of PS $\theta$ in terms of $(C,f,R,L_1,L_2)$ is achieved via \eqref{theta} for modeling IRS for efficient BD.

\subsection{RA-based IRS Circuit Characterization}
Further, we investigate RA $\mathcal{R}$ for characterizing the IRS circuitry for BD. In this regard, first we square both sides of the equations \eqref{cos}, \eqref{sin} and then add them to result in, 
\vspace{0.25cm}
\begin{equation}\label{betabw}
    \frac{(\mathcal{A} \mathcal{C} \mathcal{R})^2+(\mathcal{B} \mathcal{D}\mathcal{R})^2+(\mathcal{B}\mathcal{C}\mathcal{R})^2+(\mathcal{A}\mathcal{D} \mathcal{R})^2}{(\mathcal{C}^2\mathcal{R}^2+\mathcal{D}^2\mathcal{R}^2)^2} = 1.
  \end{equation}
Now, from \eqref{betabw}, the value of $\mathcal{R}$, is obtained as,
\vspace{0.25cm}
\begin{equation}\label{amplitude}
  \Rightarrow  \mathcal{R} = \sqrt\frac{\mathcal{A}^2+\mathcal{B}^2}{\mathcal{C}^2+\mathcal{D}^2}\triangleq \mathcal{R}(C,f,R,L_1,L_2).
\end{equation}

Hence, a closed-form expression of RA $\mathcal{R}$ in terms of $(C,f,R,L_1,L_2)$ has been achieved by \eqref{amplitude} for efficient IRS design.

\subsection{Capacitance Relationship with PS and RA}
In this Section, for IRS designer's perspective, we intend to provide the expressions for equivalent capacitance of IRS unit cell in terms of PS, RA and other physical properties of IRS. So to start with, \eqref{theta} can be first rearranged such that,
\vspace{0.25cm}
\begin{equation}\label{c_ps1}
    \tan\theta= \frac{(\mathcal{G}+\frac{Z_0}{\omega C})(\mathcal{N}+\frac{L_1}{\omega C})-(\mathcal{K}+\frac{L_1}{C})(\mathcal{M}-\frac{Z_0}{\omega C})}{(\mathcal{K}+\frac{L_1}{C})(\mathcal{N}+\frac{L_1}{C})+(\mathcal{G}+\frac{Z_0}{\omega C})(\mathcal{M}-\frac{Z_0}{\omega C})},
\end{equation}
\vspace{0.25cm} 
\noindent where $\mathcal{G}\triangleq\omega R L_1- Z_0\omega L_1-Z_0\omega L_2$,  $\mathcal{K} \triangleq -\omega^2 L_1 L_2- R Z_0$,  $\mathcal{N} \triangleq -\omega^2 L_1 L_2+ R Z_0$ and $\mathcal{M} \triangleq -\omega R L_1 +Z_0\omega L_1+ Z_0 \omega L_2$. 

Further, \eqref{c_ps1} can be simplified to get a quadratic equation in $C$ as,
\vspace{0.25cm}
\begin{equation}\label{c_ps2}
    C^2(\alpha-\psi \tan\theta)+C(\beta-\delta \tan\theta)+ (\gamma-\mu \tan\theta)=0,
\end{equation}

where $\alpha\triangleq\mathcal{G}\mathcal{N}$-$\mathcal{K}\mathcal{M}$, $\beta\triangleq\left(\mathcal{G}L_1+\frac{\mathcal{N}Z_0}{\omega}+\frac{\mathcal{K}Z_0}{\omega}-\mathcal{M}L_1 \right)$, $\gamma\triangleq\frac{2L_1 Z_0}{\omega}$, $\psi\triangleq(\mathcal{K}\mathcal{N}$+$\mathcal{G}\mathcal{M})$, $\delta$ $\triangleq$ $\left(\mathcal{K}L_1+\mathcal{N} L_1 -\frac{\mathcal{G}Z_0}{\omega}-\frac{\mathcal{M}Z_0}\omega \right)$  and $\mu\triangleq L_1^2-\left(\frac{Z_0}{\omega}\right)^2$. 

So, \eqref{c_ps2} can now be easily solved to obtain the closed-form solution of $C$, i.e. $C_\theta$ as,
\vspace{0.25cm}
\begin{equation} \label{cvstheta}
     C_\theta= \frac{-\rho+ \sqrt{(\delta \tan\theta-\beta)^2-4(\psi \tan\theta-\alpha)(\mu \tan \theta -\gamma)}}{2(\psi \tan\theta-\alpha)},
 \end{equation}
 where $\rho \hspace{-0.5mm}\triangleq \hspace{-0.5mm}\delta \tan\theta-\beta$, and $C_\theta$ is a function of $(\theta,L_1,L_2,f, R)$.

Moreover, by performing few mathematical rearrangements, \eqref{amplitude} results in,
\vspace{0.25cm}
\begin{equation} \label{c_ra1}
\mathcal{R}= \sqrt{\frac{\left(\mathcal{U}^2+\mathcal{P}^2\right)+\frac{1}{C}\left(2RL_1+\frac{2\mathcal{P}Z_0}{\omega}\right)+\frac{1}{C^2}\left(\frac{Z_0}{\omega}\right)^2}{(\mathcal{Q}^2+\mathcal{T}^2)+\frac{1}{C}\left(2QL_1-\frac{2\mathcal{T}Z_0}{\omega}+\frac{Z_0^2}{\omega}\right)+\frac{L_1^2}{C^2}}},
\end{equation}
\noindent where $\mathcal{U}$$\triangleq$$-\omega^2 L_1 L_2 -  R Z_0$, $\mathcal{P}$$\triangleq$ $\omega R L_1-Z_0 \omega L_1 - Z_0 \omega L_2$, $\mathcal{Q}$ $\triangleq$ $-\omega^2 L_1 L_2+ R Z_0$, and $\mathcal{T}$$\triangleq$ $\omega R L_1 + Z_0\omega L_1 + Z_0 \omega L_2$. Consequently, similar to \eqref{c_ps2}, \eqref{c_ra1} can be rewritten as,
\vspace{0.25cm}
 \begin{equation}\label{c_ra2}
    C^2(\chi-\lambda \mathcal{R})+C(\sigma-\Delta \mathcal{R})+ (\phi-\zeta \mathcal{R})=0,
 \end{equation}

\noindent where $\chi \triangleq \mathcal{U}^2+\mathcal{P}^2$, $\sigma\triangleq \left(2RL_1+\frac{2\mathcal{P}Z_0}{\omega}\right)$, $\phi\triangleq (\frac{Z_0}{\omega})^2$, $\lambda \triangleq \mathcal{Q}^2+\mathcal{T}^2$, $\Delta \triangleq \left(2QL_1-\frac{2\mathcal{T}Z_0}{\omega}+\frac{Z_0^2}{\omega}\right)$ and $\zeta \triangleq L_1^2$. 

Next, \eqref{c_ra2} is solved to result in the closed-form solution $C_\mathcal{R}$ of $C$ as,
\vspace{0.25cm}
\begin{equation} \label{cvsbeta}
     C_\mathcal{R} = \frac{-\varphi+ \sqrt{(\Delta \mathcal{R}^2-\sigma)^2-4(\lambda \mathcal{R}^2 -\chi)(\zeta \mathcal{R}^2 -\phi)}}{2(\lambda \mathcal{R}^2-\chi)},
 \end{equation}
\noindent where $\varphi \hspace{-0.5mm}\triangleq \hspace{-0.5mm}\Delta \mathcal{R}^2-\sigma$, and  $C_\mathcal{R}$ is a function of $(\mathcal{R}, L_1, L_2, f, R)$.

Conclusively, the closed-form relationships provided by \eqref{theta}, \eqref{amplitude}, \eqref{cvstheta} and \eqref{cvsbeta} will help in controlling the IRS BD by selecting appropriate IRS effective capacitance for desired PS and RA.   

\section{Results and Discussion}
In this Section, we provide numerical results for analyzing the proposed IRS circuit model for BD. For simulations, unless specified exclusively, we have used system parameter values as: $L_1 \in [2.3,2.5]$ nH, $L_2 \in [0.4,0.56]$ nH, $R\in [2,4]$ $\Omega$. $Z_0= 377$ $\Omega$, $C \in [0.47, 2.35]$ pF, $f=2.4$ GHz  and phase changes from $-\pi$ to $\pi$. Further, we assume practical surface mount varactor diodes, which is in accordance with device manual SMV1231-079 \cite{sky}. 

 \begin{figure}
     \centering
     \includegraphics[width=3.48in, height=2.8in]{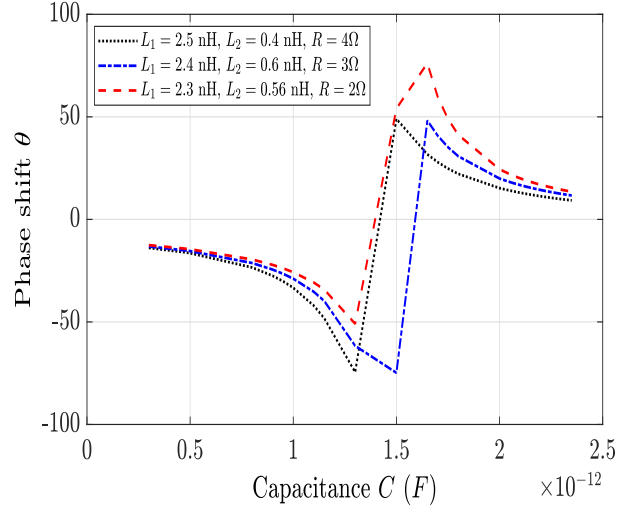}
     \caption{Effect of $C$ on PS for different electrical parameters.}
     \label{fig2}
 \end{figure}
 
 \begin{figure}
     \centering
     \includegraphics[width=3.48in,height=2.8in]{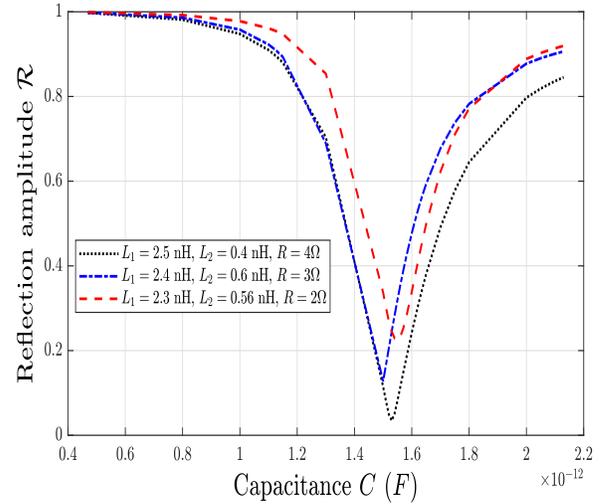}
     \caption{Impact of $C$ on RA for different electrical parameters.}
     \label{fig3}
 \end{figure}
 
 \begin{figure}
    \centering
    \includegraphics[width=3.48in,height=2.8in]{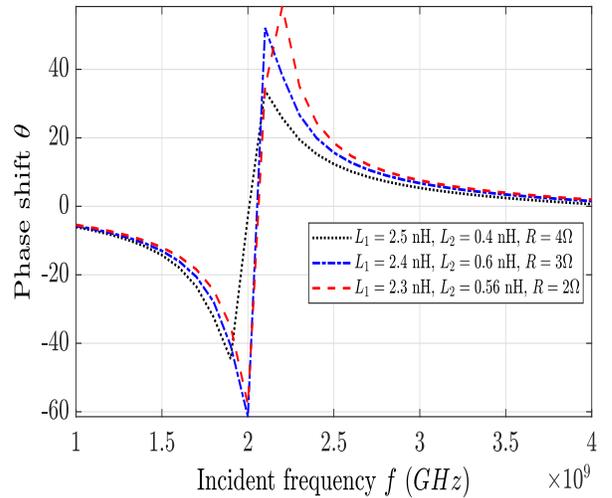}
    \caption{Effect of different IFs on PS.}
    \label{fig4}
\end{figure}

\begin{figure}
    \centering
    \includegraphics[width=3.48in,height=2.8in]{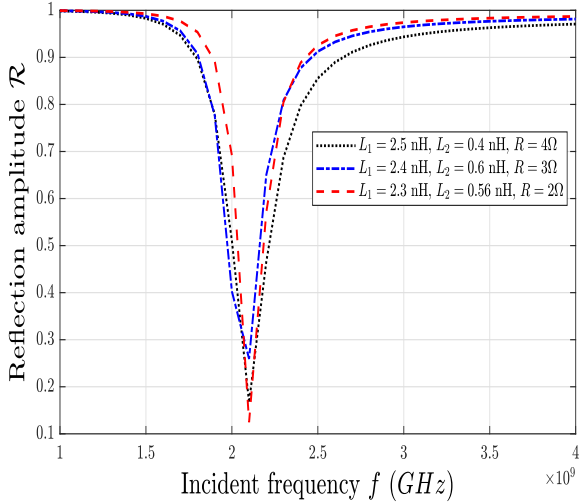}
    \caption{Influence of different IFs on RA.}
    \label{fig5}
\end{figure}

\begin{figure}
    \centering
    \includegraphics[width=3.48in,height=2.8in]{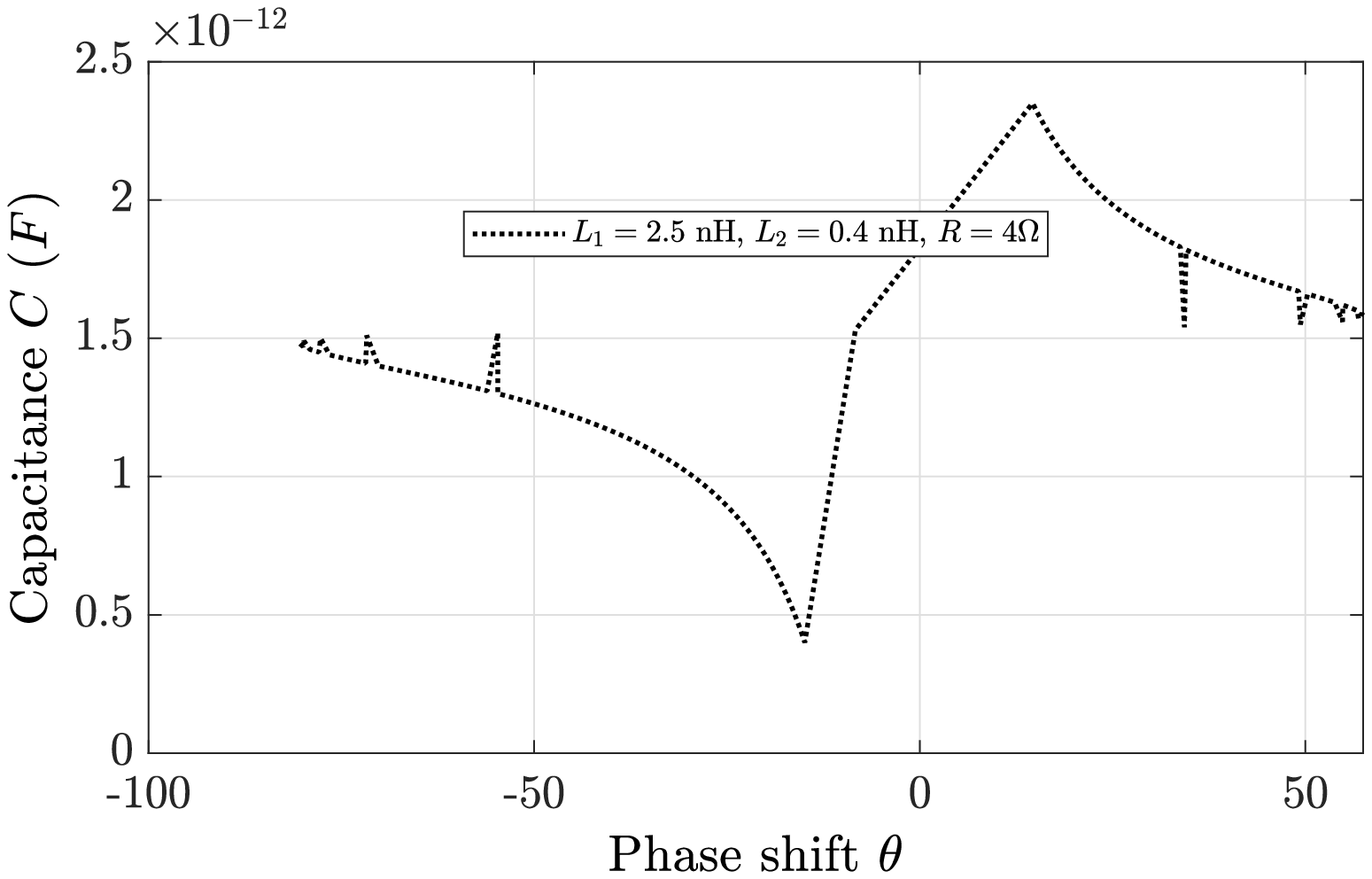}
    \caption{Variation of $C$ for different PS at receiver.}
    \label{fig6}
 \end{figure}
 
 \begin{figure}
    \centering
    \includegraphics[width=3.48in,height=2.8in]{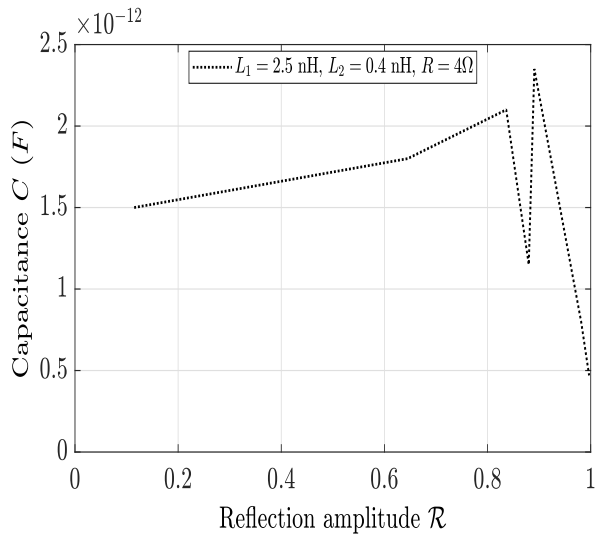}
    \caption{Variation of $C$ for different RA at receiver.}
    \label{fig7}
 \end{figure}
 
Fig. \ref{fig2} shows the impact of equivalent capacitance on PS of the IRS element by varying different electrical parameters such as $L_1, L_2, R$. All the curves are observed to be inline with \eqref{theta}. The highest value of PS has been found to be $76^{\circ}$ at $1.6$ pF with $L_1=2.3$ nH, $L_2=0.56$ nH and $R=2$ $\Omega$. Similarly, Fig. \ref {fig3}, depicts the impact of variation of $C$ on RA at the communication receiver according to \eqref{amplitude}. The minimum amplitude $\mathcal{R}=0.04$ appears at $C=1.55$ pF for $L_1=2.5$ nH, $L_2=0.4$ nH and $R=4$ $\Omega$.

The effect of transmitted signal frequency on PS of an IRS element is presented in Fig. \ref {fig4} for different sets of electrical parameters. Here, the value of $C$ is fixed at $2$ pF. It is clear that if the IF is $2.2$ GHz, then maximum PS is achieved at Rx. However for minimum PS, IF should be around $2$ GHz in order to achieve maximum PS at Rx. 

Likewise, Fig. \ref {fig5} depicts the impact of varying IF on RA. The nature of the curves is similar to Fig. \ref{fig3}, where we observe that around center frequency $2.4$ GHz, there are different RA values, $0.88,0.87,0.79$ that are obtained at $R=2$ $\Omega$, $3$ $\Omega$ and $4$ $\Omega$ respectively. Moreover, we obtain minimum amplitude $0.12$ at $f=2.1$ GHz, and maximum RA at $f=1$ GHz.

Next in Fig. \ref {fig6}, for an IRS designer to achieve high beamforming gain, our aim is to illustrate the variation of $C$, with respect to different values of PS obtained at receiver. In accordance with \eqref{cvstheta}, we observe that for given PS such as $-80^{\circ}$, $56^{\circ}$ at Rx, the value of $C$ to be set by the designer are $1.48$ pF, $2.3$ pF respectively. 

Further as per \eqref{cvsbeta}, in a similar plot Fig. \ref {fig7}, variation of $C$ has been depicted for different RAs achieved at Rx with particular set of electrical parameters, i.e. $L_1=2.5$ nH, $L_2=0.4$ nH, and $R= 4$ $\Omega$. Clearly, for no reflection or minimum RA at Rx, $C$ has to be fixed at $1.5$ pF. 

Finally in Table I, for different sets of desired $\theta$, $\mathcal{R}$ and $f$, corresponding $C$ values are presented, which are to be set by the designer at transmitter to achieve the best possible IRS-assisted beamforming gains.
\begin{table}
\centering
\caption{Range of circuit parameters for desired $\theta$, $\mathcal{R}$}
\begin{tabular}{!{\color{black}\vrule}c!{\color{black}\vrule}c!{\color{black}\vrule}c!{\color{black}\vrule}c!{\color{black}\vrule}} 
\hline
\multicolumn{3}{!{\color{black}\vrule}c!{\color{black}\vrule}}{Design Parameters} & \multirow{2}{*}{$C$ (pF)}  \\ 
\cline{1-3}
$\theta$$^{\circ}$ & $\mathcal{R}$ & $f$ (GHz)  &   \\ 
\hline
30         & 0.39      & 2.0         & 2.34   \\ 
\hline
35         & 0.30      & 2.1         & 2.13  \\ 
\hline
40         & 0.7       & 2.2        & 1.99   \\ 
\hline
45         & 0.7       & 2.3      & 1.81    \\ 
\hline
55         & 0.8       & 2.4     & 1.56    \\
\hline
\end{tabular}
\end{table}

\section{Conclusion}
In this work, we proposed a novel IRS circuit characterization model for efficient BD, which is based on PS, RA and effective IRS capacitance. We derived closed-form expressions for PS and RA in terms of signal IF and different IRS electrical parameters. We also obtained closed-form value of effective IRS capacitance as a function of PS, RA and other electrical parameters of IRS circuit. Further, we provided key design insights for model analysis, thereby also illustrating the range of $C$ required at transmitter to get desired PS and RA at receiver for high beamforming gains. This work will serve as a benchmark for practical IRS-aided BD for wireless communication, while catering for realistic values of circuit parameters which can be taken for a given transmission frequency range.

In future, we intend to utilize the proposed IRS model to efficiently to control BD for optimal resource allocation in energy harvesting based wireless multi-antenna systems.

\makeatletter
\renewenvironment{thebibliography}[1]{%
  \@xp\section\@xp*\@xp{\refname}%
  \normalfont\footnotesize\labelsep .5em\relax
  \renewcommand\theenumiv{\arabic{enumiv}}\let\p@enumiv\@empty
  \vspace*{-1pt}
  \list{\@biblabel{\theenumiv}}{\settowidth\labelwidth{\@biblabel{#1}}%
    \leftmargin\labelwidth \advance\leftmargin\labelsep
    \usecounter{enumiv}}%
  \sloppy \clubpenalty\@M \widowpenalty\clubpenalty
  \sfcode`\.=\@m
}{%
  \def\@noitemerr{\@latex@warning{Empty `thebibliography' environment}}%
  \endlist
}
\makeatother

\bibliographystyle{IEEEtran}
\bibliography{reference}
\end{document}